\documentclass[twocolumn,showpacs,preprintnumbers,amsmath,amssymb]{revtex4}

\usepackage{amsmath}
\usepackage{epsfig}
\usepackage{epsf,afterpage}
\usepackage{amssymb}
\newcommand{\be}{\begin{equation}}
\newcommand{\ee}{\end{equation}}

\usepackage{graphicx}
\usepackage{dcolumn}
\usepackage{bm}

\begin{document}

\title{On the photon polarization
in radiative $B\to \varphi K \gamma$ decay.}

\author{V.D.Orlovsky, V.I.Shevchenko}
\affiliation{%
Institute of Theoretical and Experimental Physics
\\B.Cheremushkinskaya 25, 117218 Moscow, Russia
}%

\date{\today}

\begin{abstract}
The photon polarization in radiative decays $B\to Y \gamma$ is known
to be a subtle probe of the effective Lagrangian structure and
possible New Physics effects. We discuss exclusive decay mode
$B^-\to \varphi K^- \gamma$ where the experimentally distinct final
state makes analysis especially promising. The possibility to
extract information on the photon polarization out of the data
entirely depends on the partial waves interference pattern in the
$\varphi K^-$ system.
\end{abstract}

\maketitle

\section{Introduction}

The main efforts of modern high energy physics are devoted to the
searches of phenomena beyond the Standard Model (SM) and
correspondingly to constraining of different SM extensions. Flavor
physics is an important area of this activity: $B$, $D$ and
$K$-meson decays studies have brought a lot of information about
different aspects of CKM paradigm and suggest promising places to
look for New Physics (NP). The unique experimental opportunities for
the $b$-physics part of this research programme have been related to
BaBar and Belle experiments, while the main hope for now is
concentrated on LHCb experiment at CERN with prospects for Super-B
factory as a possible future project.

Among a wide variety of rare decays radiative $B$-meson decays $B
\rightarrow Y \gamma$ are especially distinctive (and sometimes even
called "the standard candles" of flavor physics \cite{neubert}). The
first obvious reason is that electromagnetic part of these decays is
under full theoretical control, while from experimental point of
view the energetic photon serves as a clean and unambiguous decay
signal. This allowed to develop effective theory of such decays and
also to obtain impressive experimental data on the corresponding
branching ratios (see, e.g. \cite{ali} and references therein).

Unfortunately comparison of experimentally measured branching ratios
with theoretical predictions is plagued by hadron uncertainties of
the latter. This motivates constant interest to theoretical and
experimental studies of "gold-plated" observables, unaffected by
hadron uncertainties. Radiative decays provide polarization pattern
of emitted photons (corresponding to angular correlations in the
final hadron state) as a good example of such observable.

Moreover, it was argued in \cite{AGS} that measurements of the
photon polarization in the final state turn out to be an effective
tool for the NP searches. The point is that photons, emitted in the
$B^-$ and $\bar{B}^0$-meson decays are predominantly left-handed
(and right-handed for the $B^+$ and $B^0$ decays) in the SM, while
the admixture of photons with "wrong"\, polarization may be rather
large in some SM extensions like e.g. Left Right Symmetric Model
(LRSM) or MSSM (Minimal Supersymmetric Standard Model). The
information one can get in this way is extremely interesting since
it provides a typical example of what is known as the "null tests"
of the SM \cite{NullTests}. It probes internal Lorentz structure of
photon emission vertex and hence essential features of the effective
Hamiltonian structure.

There have been suggested several ways to look for signals beyond
the SM through the photon helicity tests. In particular, the
admixture of right-handed photons may be found via the
time-dependent CP-asymmetry in $B^0(t) \rightarrow f^{CP} \gamma$
decays, where $f^{CP} = K^{*0} \rightarrow K_S \pi^0$:
\begin{eqnarray}
{\cal A}(t) = \frac{\Gamma(B^0(t)\rightarrow f^{CP}) -
\Gamma(\bar{B}^0(t)\rightarrow f^{CP})}{\Gamma(B^0(t)\rightarrow
f^{CP}) + \Gamma(\bar{B}^0(t)\rightarrow f^{CP})} \nonumber \\
=S\sin(\Delta
m_B t) - C\cos(\Delta m_B t).
\end{eqnarray}
The mixing-induced asymmetry $S$ is proportional to the $A_R/A_L$
ratio of the polarization amplitudes, which corresponds to right-
and left-handed photon emission and expected to be less than a few
percent (see below) in the SM \cite{AGS,NullTests,AGSH}.

Another method makes use the photons from the $B \rightarrow (K^*
\rightarrow K \pi) \gamma$ decay, converting in the detector
material into the electron-positron pair (see recent paper \cite{c}
in this respect). For these processes the distribution in the angle
$\phi$ between $e^+e^-$ and $K \pi$ planes should be isotropic for
purely circular polarization, the deviations from this isotropy
depends on the same parameter $A_R/A_L$, indicating the presence of
right-handed photons \cite{e+-1,e+-2,e+-3,e+-4}. So, the angular
distribution for real photons is given by
\begin{equation}
\frac{d\sigma}{d\phi} \propto 1 + \eta \frac{A_LA_R}{A_L^2+A_R^2}
\cos(2\phi + \delta'),
\end{equation}
where $\eta$ and $\delta'$ are some hadronic parameters of no
importance for us here.

Alternatively, one can study baryon decays $\Lambda_b \rightarrow
\Lambda \gamma \rightarrow p \pi \gamma$ and measure the photon
polarization directly. It is proportional here to the
forward-backward asymmetry of the proton with respect to $\Lambda_b$
in the rest frame of $\Lambda$ or related to $\Lambda_b$
polarization and forward-backward asymmetry of $\Lambda$ momentum
for the polarized $\Lambda_b$'s \cite{lambda1,lambda2,lambda3}.

In this paper we follow the standard method, which makes use of
angular correlations among the three-body decay products in $B
\rightarrow P_1 P_2 P_3 \gamma$, where $P_i$ are either pions or
kaons. This technique was suggested in \cite{Gronauold,Gronaubase}
and used for the decay $B \rightarrow K \pi \pi \gamma$ with the
manifest summation over intermediate hadron resonances. We consider
the radiative decay mode $B \rightarrow (\varphi\rightarrow K^+K^-)
K \gamma$ in the present paper. The mode $B \rightarrow \varphi K
\gamma$ is rather distinctive with many desirable features from the
experimental point of view: the finite state is a photon plus only
charged mesons (for charged $B$-mesons), the fact that $\varphi$ is
narrow reduces the effects of intermediate resonances interference
etc. The branching fraction for this decay mode was measured by
BaBar and Belle collaborations \cite{Belle,Babar}: \be {\cal
B}(B^-\rightarrow \varphi K^- \gamma) = (3.5\pm 0.6)\times
10^{-6}\ee and this decay channel is currently being studied under
LHCb rare decays program.

The general qualitative physical picture behind the photon
polarization measurement procedure discussed in the present paper
can be explained as follows. The $b$-quark belonging to the initial
pseudoscalar $B$ meson decays due to weak penguin process into a
photon $\gamma$ and $s$-quark. The latter forms the hadron system
$Y$ (together with the spectator), which is characterized by total
angular momentum $J \ge 1$ and its projection $\lambda$. Strong
dynamics causes consequent decay of $Y$ into a pseudoscalar $P_3$
(where the spectator quark goes) and a vector or tensor $T$ (where
the $s$-quark goes). \be Y(J^P, \lambda)  \to P_3 [T\to P_1 P_2] \ee
We have ${\vec J} = {\vec j}_T + {\vec l}$ where ${\vec l}$ is
relative orbital momentum of the states $T$ and $P_3$. The tensor
helicity $\lambda_T$ carries information about the $s$-quark
helicity, which in turn is correlated with the photon polarization.
The partial wave amplitude takes the form: \be
 A_{l\lambda} \propto \sum\limits_{\lambda_T = -j_T}^{j_T}
( l,0;j_T,\lambda_T|J,\lambda_T) \cdot {\bar A}_{\lambda_T \lambda}
\ee where  $( l,0;j_T,\lambda_T|J,\lambda_T )$ are Clebsch-Gordan
coefficients. If relative angular momentum between $P_3$ and $T$ is
zero we have no way to uncover this information since for $l=0$ all
polarization states of $T$ enter on equal footing and the amplitude
$A_{0\lambda} $ has no sensitivity to $\lambda$. But it is not the
case if $l\neq 0 $ and then nontrivial asymmetric interference
pattern \be |A_R|^2 - |A_L|^2 \propto {\vec p}_\gamma \cdot [{\vec
p}_1 \times {\vec p}_2 ] \label{poi} \ee starts to show up. This
picture is applicable to both resonant amplitudes in the $B\to
K\pi\pi\gamma$ channel and non-resonant $B\to K\varphi\gamma$
channel (where $\varphi$ plays the role of $T$).

It is convenient to define the total decay amplitude as a
convolution of weak radiative amplitude $c_{L,R} = A(B \to Y
\gamma_{L,R})$ and strong polarization amplitude $A_{L,R} = A(Y\to
[\varphi \to KK] K)$ for the consequent decay
 corresponding to the left-
and right-polarized resonance $Y$, respectively (including all
necessary form-factors and Breit-Wigner forms).
 The
photon polarization parameter $\lambda_\gamma^{(i)}$ defined in
terms of amplitudes ratio for the decay $B \rightarrow Y^{(i)}
\gamma_{(L,R)}$ could depend on the final state $Y^{(i)}$ quantum
numbers. However, due to parity conservation by the strong
interactions it does not \cite{Gronauold}. Moreover, since we
consider in what follows $[\varphi K]$ system in a state with the
fixed quantum numbers, we can define the photon polarization
parameter simply as
\begin{equation}\label{lambda2}
\lambda_\gamma = \frac{|c_R|^2-|c_L|^2}{|c_R|^2+|c_L|^2}
\end{equation}
Another general comment is worth making. According to the standard
quantum mechanics, the expression for partial branching ratio
contains sum over final states, which in our case is a state of
hadronic system plus a photon of definite helicity. From general
principles it is clear however that the amplitudes, corresponding to
emission of left-handed and right-handed photons do not interfere
since they correspond to different final states and, as a matter of
principle, the photon helicity can be measured independently (for
example in gedanken way by measuring angular momentum of the
detector). As a result, general expression for the partial decay
width takes the following general form \cite{Gronauold,Gronaubase}:
\be \frac{d\Gamma}{d\Phi} \propto |c_L A_L|^2 + |c_R A_R|^2
\label{uuu} \ee where $d\Phi$ is the final particles phase space
(see exact form after eq.(\ref{terf})) and the polarization
amplitudes ${A}_R$ (${A}_L$) correspond to the left-(right)-handed
photon emission. There is no interference terms $\sim
|{A}_L^*{A}_R|$ in the expression (\ref{dgamma}) and this fact is
completely independent on the structure of the amplitudes (i.e.
whether they are real or complex, presence or absence of NP effects
etc). This is in contrast with results of the recent paper
\cite{Sonibase}, where the same decay mode $B\to K\varphi\gamma$ is
considered.

Taking into account the definition (\ref{lambda2}) one has for the
partial decay width
\begin{equation}\label{dgamma}
\frac{d\Gamma}{d\Phi} \propto |{A}_R|^2+ |{A}_L|^2 + \lambda_\gamma
(|{A}_R|^2 - |{A}_L|^2)
\end{equation}
To find $\lambda_\gamma $ one has to extracts from the branching
ratio (\ref{dgamma}) the angular part (\ref{poi}), sensitive to the
discussed asymmetry.

\section{Photon polarization in the Standard Model and beyond}

In the SM the radiative decay of $b$-quark is governed by the lowest
order effective Hamiltonian
\begin{eqnarray}
H_{eff}=-\frac{4 G_F}{\sqrt{2}} V_{tb} V^*_{ts} (C_{7R} O_{7R} +
C_{7L} O_{7L} ) \nonumber \\
 O_{7L,R} = \frac{em_b}{16\pi^2}F_{\mu\nu} \bar{s}
\sigma^{\mu\nu} \frac{1\pm\gamma_5}{2}b  \label{ham}
\end{eqnarray}
Here $C_{7L,R}$ are the Wilson coefficients corresponding to the
amplitude for emission of left or right handed photons in the $b_R
\rightarrow q_L \gamma_L (b_L \rightarrow q_R \gamma_R)$ decays.
This can be seen by representing the electromagnetic field tensor
for left-(right-) polarized photons: $F^{L,R}_{\mu\nu} =
\frac{1}{2}(F_{\mu\nu}\pm i\tilde{F}_{\mu\nu})$, where
$\tilde{F}_{\mu\nu} = \frac{1}{2}\varepsilon_{\mu\nu\sigma\rho}
F^{\sigma\rho}$. Using the identity $\sigma_{\mu\nu}\gamma_5 =
\frac{i}{2}\varepsilon_{\mu\nu\alpha\beta}\sigma^{\alpha\beta}$ one
can see that only $F^L_{\mu\nu}$ survives in the first term of
(\ref{ham}) and only $F^R_{\mu\nu}$ in the second one. In the SM the
ratio, measuring the part of "wrong"\, helicity photons
$|C_{7R}/C_{7L}| \propto m_s/m_b$, since only the left-handed
components of the external fermions couple to $W$-boson. However,
besides the kinematical corrections controlled by the mass ratio
$m_s/m_b$ there are also QCD corrections - perturbative and
nonperturbative. They were estimated as sufficiently large - about
10\% - in the papers \cite{Lambda/mb1}, \cite{Lambda/mb2}. More
detailed calculations, taking into account effects due to hard gluon
emission, estimate the corrections at the 3-4\% level \cite{pQCD}.
The nonperturbative corrections resulted from the soft gluon
emission by $c$-quark loop in the effective operator $O_2$ turn out
to be about $1\%$, while nonperturbative contributions from the
annihilation diagrams and other operators are of the same order or
smaller, as was estimated in the detailed light-cone sum rule method
calculations \cite{BallZwicky}. Thus the total deviation of the
right to left photons ratio from zero not exceeding 5\% in the SM
seems to be based on rather solid theoretical grounds. Larger
values, if observed, have to be interpreted as a manifestation of
NP.

The decay process $B\to Y \gamma$ receives, besides short-distance
contributions described by (\ref{ham}), also long-distance
contributions. The structure and relative role of the latter is
rather complex and was analyzed in details in \cite{gp}. There are
two outcomes of this analysis to be mentioned here. First, the
short-distance term is always leading, despite the relative
magnitude of the long-distance contributions can be sizeable.
Second, and this is of prime importance for us, the dominant
left(right)-handedness of the emitted photon is not  affected by the
long-distance terms, in other words, the long-distance amplitudes
for emission of the photon with the "wrong" polarization obey the
same hierarchy with respect the the "right" ones, as short-distance
terms do. Since we concentrate in what follows on angular
distributions and do not pretend to compute the absolute values of
the branching ratios, we can safely assume that our strong
amplitudes include both short and long distance contributions.

Summarizing the discussion above performing the fitting procedure
for particular partial decay width of $B^-$ meson with (\ref{uuu})
one has to obtain $\lambda_\gamma = -1 + \xi^2$ in the SM, where the
factor $\xi$ not exceeding 3-5 \% takes into account all possible SM
corrections for right-handed photons admixture. On the other hand,
there are NP scenarios, where the suppression of "wrong" \, helicity
photon emission is absent. A good example are left-right symmetric
models, in which the enhancement of the right-handed photons
fraction is due to the $W_L - W_R$ mixing, and chirality flip along
the internal $t$-quark line in the loop leads to large factor
$m_t/m_b$ in the amplitude for producing right-handed photons. The
predictions for the mixing-induced CP asymmetry in $B^0\rightarrow
K^*\gamma$ under assumption that the radiative decay rates agree
with the SM expectations are \cite{AGS}
\begin{equation}
A(t) \approx \mp
2\cdot(120\zeta)\cdot\sqrt{1-(120\zeta)^2}\cos(2\beta)\sin(\Delta
m_Bt),
\end{equation}
where $10^\circ < \beta < 35^\circ$ and the mixing parameter $\zeta$
is constrained by experimental observations $\zeta \leq 3\cdot
10^{-3}$, so the asymmetry can be as large as 50\%. It was shown
that within the unconstrained minimal supersymmetric SM (uMSSM)
strong enhancement of order $m_{\tilde{g}}/m_b$ is possible due to
chirality flip along the gluino line and left-right squark mixing.
In this case the parameter $\lambda_\gamma$ can take any value
between $-1$ and $1$ \cite{MSSM}. The model with anomalous
right-handed top couplings \cite{anom_top} predicts sizeable
contributions in $A_R$, resulted in the polarization parameter $-1 <
\lambda_\gamma \lesssim -0.12$. In models with non-supersymmetric
extra dimensions there are also no reasons for right-handed photon
to be suppressed with respect to the left-handed one, so that
$\lambda_\gamma$ is close to zero and mixing-induced CP asymmetries
are of the order of one \cite{extra_dim}.

\section{Angular distribution in the $B\to [\varphi K]^1 \gamma$  decay}

As is well known in studies of many-body sequential decays one can
use either "helicity"\, or "tensor"\, formalism. Since our interest
is focused on angular dependencies, the former approach is most
suitable \cite{helicity}, see \cite{Chung} for introduction and
further references. An amplitude for the two-body decay
$Y\rightarrow 1+2$ of the resonance of spin-parity $J^P$ with the
$z$-component $M$ into particles 1 and 2 with spins and helicities
$s_1, \lambda_1$ and $s_2, \lambda_2$, respectively is given in
terms of finite rotation of the $z$-axis to the axis of $Y$-decay:
\begin{equation}
A(Y\rightarrow 1 + 2) = N_JA_{\lambda_1\lambda_2}^J
D^{J*}_{M\lambda}(\phi,\theta,0),
\end{equation}
where $\lambda = \lambda_1 - \lambda_2$, the spherical angles
$(\theta, \phi)$ define the direction of the particle 1 momentum
relative to the $z$-axis. All angular dependence is concentrated in
the standard rotation matrix $D^j_{mm'}(\alpha,\beta,\gamma)$
\begin{eqnarray}
D_{mm'}^j(\alpha,\beta,\gamma) = e^{-im\alpha}d^j_{mm'}(\beta)
e^{-im\gamma} \nonumber \\
 d^j_{mm'}(\beta) = \langle jm|e^{-i\beta
J_y}|jm' \rangle.
\end{eqnarray}

Let us define the coordinate systems and angles, related to the
decay of interest. The $z$ axis in the $[\varphi K]$ rest frame is
anti-parallel to the photon momentum: ${\bf p}_\gamma/|{\bf
p}_\gamma| = -{\bf e}_z$. There is a plane defined by the 3-momenta
of final state kaons \be B\to [\varphi (-{\bf p}_3)\> K({\bf p}_3)]
\gamma  \to K({\bf p}_1)\> K({\bf p}_2)\> K({\bf p}_3) \gamma \ee in
the $[\varphi K]$ rest frame ${\bf p}_1 + {\bf p}_2 + {\bf p}_3 =
{\bf 0}$. We define the $z'$ axis as being orthogonal to this plane,
while $y'$ axis is directed along ${\bf p}_3$: \be {\bf e}_{z'} =
[{\bf p}_1 \times {\bf p}_2]/|[{\bf p}_1 \times {\bf p}_2]| ; \;\;\;
{\bf e}_{y'} = {\bf p}_3 / |{\bf p}_3| \ee and ${\bf e}_{x'} = [{\bf
e}_{y'}\times {\bf e}_{z'}] $. Consequently, it is convenient to
define the corresponding angles. First, we define the angle $\theta$
between the $z$ and $z'$ axes, i.e. between the photon momentum and
normal to the $[\varphi K]$ decay plane. Second, there are polar and
azimuthal angles $(\eta, \phi)$ for the vector ${\bf p}_3$ in the
$(x,y,z)$ frame. Note that these angles are defined in the $[\varphi
K]$ rest frame and the angle $\phi$ is unobservable. In analogous
way in the $\varphi$ rest frame one has $\varphi({\bf p} = {\bf 0})
\to K({\bf p}_1^*) K(- {\bf p}_1^*)$ and the polar angle $\theta^*$
of the vector ${\bf p}_1^*$ is defined with respect to $y'$ axis,
while the azimuthal angle $\phi^*$ measures the rotations of ${\bf
p}_1^*$ around this axis. It is not independent and can be simply
expressed as a function of $\eta$ and $\theta$. To summarize, we
have \begin{eqnarray} \cos\theta = ({\bf e}_{z}\cdot {\bf e}_{z'})
\;\;\;\;
\cos\eta = ({\bf e}_{z}\cdot {\bf e}_{y'})  \nonumber \\
 \cos\theta^* = ({\bf e}_{y'} \cdot {\bf p}_{1}^*) / |{\bf
p}_{1}^*| \;\;\;\; \sin\phi^* = \cos\theta / \sin\eta \end{eqnarray}
where ${\bf p}_{1}^*$ is the momentum of the first (taken as e.g.
the fastest) kaon resulted from $\varphi$ decay in the $\varphi$
rest frame. Figure 1 represents our momenta and angle conventions.

The amplitude $A_M$ (where $M=1$ corresponds to the right-handed
photon and $M=-1$ to the left-handed one) for the sequential decay
$Y\equiv[\varphi K]\rightarrow \{\varphi\rightarrow K({\bf p}_{1})\>
K({\bf p}_{2})\} \> K({\bf p}_{3})$ is proportional to the standard
convolution:
\begin{widetext}
\begin{equation}\label{am}
A_M \propto \sum_{J=1,2,.. \atop \lambda_\varphi=0,\pm 1}
\left\langle K^+({\bf p}_1)K^-({\bf p}_2)|\Delta H_\varphi |
\varphi(-{\bf p}_{3}, \lambda_\varphi) \right\rangle \left\langle
\varphi(-{\bf p}_{3}, \lambda_\varphi) K^-({\bf p}_3)| \Delta H_Y |
Y^-({\bf 0}; J^PM) \right\rangle
\end{equation}
\end{widetext}
The first factor is the standard expression for $p$-wave decay of
the vector $\varphi$-resonance into two $K$-mesons: \begin{eqnarray}
\left\langle K^+({\bf p}_1)K^-({\bf p}_2)|\Delta H_\varphi |
\varphi(-{\bf p}_3, \lambda_\varphi) \right\rangle = \nonumber
\\ = \bar{a}_p \cdot D^{1*}_{\lambda_\varphi 0}(\phi^*,\pi - \theta^*,0).
\end{eqnarray}
 The second factor in the r.h.s. of (\ref{am}) can be expanded into
the sum over the partial waves with each partial wave amplitude
$a_l$ entering with the factor
\be (2l+1)^{1/2}\cdot (l,0,1,\lambda_\varphi|J,\lambda_\varphi)\cdot
D^{J*}_{M \lambda_\varphi}(\phi,\pi-\eta,0) \label{pw} \ee It
describes the transition of the initial hadronic system at rest $Y$
of spin $J$ with $z$-component $M$, created after the photon
emission in $B\to Y\gamma$ decay into a system of $K$ and $\varphi$
mesons with definite momenta ${\bf p}_3$ and $-{\bf p}_3$,
respectively, and helicity $\lambda_\varphi = 0, \pm 1$ for
$\varphi$-resonance.

It is not known {\it a priori} how many contributions are important
in the sum over $J$ in (\ref{am}). Neither it is known how the
partial waves expansion saturates the sum (\ref{pw}). Contrary to
the case of $K\pi\pi$ channel studied in \cite{Gronaubase}, we have
here no independent information about the relative partial waves
phases. However, one has no reasons to expect strong coupling with
the closest physical state in $[\varphi K]$ channel above the
threshold $K_2(1770) \; J^P = 2^-$ since the latter dominantly
couples to $K \pi \pi$ mode. Because of experimental kinematical cut
on the maximum photon transverse momentum one is confined to the
region of not too large invariant masses of $Y$. Therefore it seems
reasonable to consider as the first approximation the simplest case
with the only $J=1$ term kept\footnote{This corresponds to the "toy
model" of \cite{Sonibase}.} in the sum (\ref{am}). We sum over both
parities of the intermediate state $Y$, which correspond to
inclusion of $s-$ and $d-$ waves for $J^P = 1^+$ state and $p-$ wave
for $J^P = 1^-$ state. Then summing over the intermediate
$\varphi$-resonance polarizations and using the explicit expressions
for $D$-functions, we obtain the differential decay rate in the
following form: $$ \frac{d\Gamma}{d\Phi} \propto \left[ c_1
\sin^2\theta^* (\cos^2\eta + \cos^2\phi^* \sin^2\eta) + \right. $$
$$ \left. + c_2 \sin^2\theta^*(\cos^2\eta + \sin^2\phi^* \sin^2\eta) +
\right.
$$ $$ + \left. c_3 \cos^2\theta^*\sin^2\eta + c_4 \sin 2\phi^*
\sin^2\theta^*\sin^2\eta + \right. $$ $$ + \left. \frac12
\sin2\theta^* \sin 2\eta (c_5 \cos\phi^* + c_6 \sin\phi^*) + \right.
$$ $$ + \left. \lambda_\gamma \left( c_7\sin^2\theta^* \cos\eta +
\right. \right. $$ \be \left. \left.+ \sin 2\theta^* \sin\eta (c_8
\cos\phi^* + c_9 \sin\phi^*)\right)\right]
 \label{terf} \ee where
the phase-space volume is determined by the integration over the
$m_{12}^2 = (p_1+p_2)^2$ and four angles $ \theta^*, \phi^*, \eta$
and unobservable angle $\phi$: $ d\Phi = dm_{12}^2 d\cos\eta d\phi
d\cos\theta^*d\phi^*$. The terms proportional to $~\cos m\eta,\>
\sin m\eta$ for $m=0$ and $m=2$ have no sensitivity to the sign of
$\lambda_\gamma $, while those for $m=1$ do have and the
contribution of these asymmetric terms to $d\Gamma$ is controlled by
the hadron parameters $c_7, c_8, c_9$.

The notation goes as follows. The partial amplitudes ratios are
given by $a_1/a_0 = r_1\exp(i\delta_1)$, $a_2/a_0 =
r_2\exp(i\delta_2)$. The coefficients read:
\begin{gather}
c_1 = r_1^2 \quad c_2 = 1+ \frac{r_2^2}{2} + \sqrt{2}r_2\cos\delta_2
\nonumber \\
c_3 = 1+ 2r_2^2 - 2\sqrt{2}r_2\cos\delta \quad  c_4 = -\sqrt\frac32
r_1 \bar\zeta_{+} \nonumber \\
c_5 = 1 - r_2^2 -\frac{r_2}{\sqrt{2}}\cos\delta_2 \quad c_6 =
 \sqrt\frac32
r_1 \bar\zeta_{-} \nonumber \\
c_7 =  \sqrt{6} r_1 \zeta_{+} \quad c_8 = \sqrt{6} r_1 \zeta_- \quad
c_9 =  \frac{3r_2}{\sqrt{2}}\sin\delta_2 \nonumber
\end{gather}
with \begin{gather} \zeta_{\pm} = \cos\delta_1 \pm (\sqrt{2})^{\mp
1} r_2 \cos(\delta_1 - \delta_2) \nonumber \\
{\bar{\zeta}}_{\pm} = \sin\delta_1 \pm (\sqrt{2})^{\mp 1} r_2
\sin(\delta_1 - \delta_2) \nonumber
\end{gather}

 If we confine ourselves by the contribution of $J^P=1^+$
states only, the result simplifies considerably, since
$c_{1,4,6,7,8}=0$ in this case and the only remaining asymmetric
term takes the form: \be |A_R|^2 - |A_L|^2 \propto c_9 \sin
2\theta^* \cos\theta \ee On the other hand, the $p-$wave
contribution  alone does not produce any asymmetry as can be deduced
from general $P$-parity arguments and directly seen from
(\ref{terf}), having no sensitivity to the photon polarization in
this case.

The expression (\ref{terf}) is the main result of this paper. In
principle one has nine independent angular structures and five
unknowns for analysis ($r_{1,2}, \delta_{1,2}$ $\lambda_\gamma$). As
a matter of principle it is perhaps more advantageous to fit unknown
strong parameters $r_{1,2}$ and $\delta_{1,2}$ from the first six
symmetric terms and then use the results to extract $\lambda_\gamma$
from the last term. Alternative practical way is to perform
integration $\int d\Gamma$ over some region of the Dalitz plot, as
suggested in \cite{Gronaubase}. It is seen however that the
possibility to proceed this way strongly depends on the actual value
of the corresponding parameters. In particular, if $p$-wave
contribution is small and also $r_2\sin\delta_2 \ll 1 $, the
discussed asymmetry will escape the detection.

Alternatively, one can try to fit the full differential rate over
the maximal available part of the final state phase space, using as
a cross-checks different constraints in the form of sum rules the
coefficients $c_1, ..,  c_9$ have to obey. Strong violation of such
sum rules would indicate importance of higher momenta terms in
(\ref{am}).

\section{Conclusion}

We have applied the general method \cite{Gronaubase} of photon
polarization parameter $\lambda_\gamma$ measurement to the radiative
$3+1$-body decay $B^-\rightarrow (\varphi\rightarrow
K^+K^-)K^-\gamma$. Such measurement is among hot topics of LHCb rare
decays physics program, and the detailed sensitivity studies are now
in progress. The only chance for this method to provide sound
experimental information on the photon polarization pattern is
strong interference between the partial waves in the $[\varphi
K]$-system with the latter being in the vector state. In fact, it is
straightforward to proceed with more general calculations, taking
into account higher momenta in $[\varphi K]$ system. Leaving aside
the cumbersome form of the results obtained, the interference
pattern becomes so complicated in this case that any reasonable
fitting procedure will certainly be impossible if higher momenta are
indeed important in the decay of interest. Thus the method is rather
restrictive from the parameter space point of view. On the other
hand, if the approximations used will happen to be correct, the
corresponding strong parameters can be determined using the same
decay mode after the LHCb data will become available. The branching
fraction for this mode is measured by BaBar and Belle collaborations
and this decay channel seems to be very promising for LHCb -- it is
expected, that one full year of LHC operation will give about 7000
selected $B\rightarrow \varphi K \gamma$ events,\footnote{Assuming
the same reconstruction efficiency as for $B\to K^* \gamma$
channel.} to be compared with only $\sim 230$ events obtained by
Belle by the end 2008.

\begin{acknowledgements}

The authors thank A.Golutvin and T.Nakada for useful discussions.
The work is supported by the grant INTAS-05-103-7571 and partly by
the grant for support of scientific schools NS-843.2006.2; contract
02.445.11.7424/2006-112. One of the authors (V.Sh.) acknowledges
support by INTAS-CERN fellowship 06-1000014-6576.
\end{acknowledgements}

\newpage

\begin{figure}[!t]
\begin{center}
\epsfxsize=10cm \epsfbox{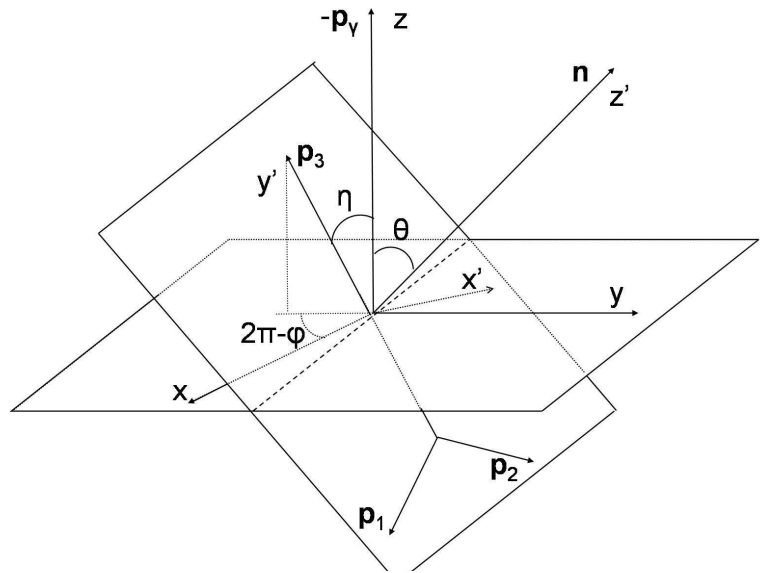} \caption{Angle conventions for
the decay $B^-\rightarrow [\varphi\rightarrow K^+({\bf p}_1)
K^-({\bf p}_2)] K^- ({\bf p}_3) \gamma({\bf p_\gamma})$.}
\end{center}
\end{figure}

\end{document}